\documentstyle[amstex,preprint,aps]{revtex}

\begin{document}
\title{Theoretical description of the ferromagnetic $\pi $-junctions near the
critical temperature}
\author{A. Buzdin and I. Baladi\'{e}}
\address{Condensed Matter Theory Group, CPMOH, UMR 5798, Universit\'{e} Bordeaux 1,\\
33405 Talence Cedex,\\
France}
\date{02/12/2002}
\maketitle

\begin{abstract}
The theory of ferromagnetic $\pi $-junction near the critical temperature is
presented. It is demonstrated that in the dirty limit the modified Usadel
equation adequately describes the\ proximity effect in ferromagnets. To
provide the description of an experimentally relevant situation,
oscillations of the Josephson critical current are calculated as a function
of ferromagnetic layer thickness for different transparencies of the
superconductor-ferromagnet interfaces.
\end{abstract}

\section{Introduction}

The strong exchange field acting on the electrons in a ferromagnet provokes
the damped oscillatory behavior of the superconducting order parameter. This
effect is at the origin of the $\pi $-junction realization in\
superconductor-ferromagnet-superconductor (S/F/S) systems \cite
{Panyukov82,Kuprianov91}. The study of the superconducting $\pi $-state
sheds more light on the coexistence of superconductivity and magnetism in
general and may be also important for superconducting electronics (see for
example \cite{Ioffe}).

Recently such $\pi $-junctions have been successfully fabricated and
experimentally studied in \cite{ryazanov,Kontos2,Blum}. While the existing
theoretical approach \cite{Kuprianov91} provides a good qualitative
description of the observed damped oscillations of the critical current as a
function of ferromagnetic layer thickness \cite{Kontos2,Blum}, a more
detailed description\ is needed to take into account the finite S/F boundary
transparency. Indeed, in \cite{Kontos2} Nb/Al/Al$_{2}$O$_{3}$/PdNi/Nb
junctions have been studied. In such junctions, the presence of the Al$_{2}$O%
$_{3}$ barrier can be modeled by a very low transparency of one of the S/F
interfaces.

In this work we develop a theory of ferromagnetic junctions with different
transparencies of the interfaces providing a general description of such
junctions near the critical temperature.

\section{General formalism}

We will concentrate on the studies of the properties of an S/F/S junction
with a thin F-layer of thickness $d$ and large superconducting electrodes$,$
see Fig. $1$. The very convenient set of equations describing an
inhomogeneous superconductivity has been elaborated by Eilenberger \cite
{Eilenberger}. These are transport-like equations for the energy-integrated
Green's functions $f$ and $g$, assuming that relevant length scales are much
larger than atomic length scales. For our geometry all quantities depend
only on one coordinate $x${\bf ,} and in the presence of the ferromagnetic
exchange field $h(x)$ acting on the electron spins in the F-layer, the
Eilenberger equations take the form (see for example \cite{L. N. Bulaevskii}%
) 
\begin{gather}
\left( \omega +ih(x)+\frac{1}{2\tau }G(x,\omega )\right) f\left( x,\theta
,\omega \right) +\frac{1}{2}v_{F}\cos \theta \frac{\partial f\left( x,\theta
,\omega \right) }{\partial x}=\left( \Delta (x)+\frac{1}{2\tau }F(x,\omega
)\right) g\left( x,\theta ,\omega \right) ,  \nonumber \\
G(x,\omega )=\int \frac{d\Omega }{4\pi }g\left( x,\theta ,\omega \right)
,\quad F(x,\omega )=\int \frac{d\Omega }{4\pi }f\left( x,\theta ,\omega
\right) ,  \label{eilenberger complet} \\
f\left( x,\theta ,\omega \right) f^{+}\left( x,\theta ,\omega \right)
+g^{2}\left( x,\theta ,\omega \right) =1,  \nonumber
\end{gather}
where $\omega =2\pi T\left( n+1/2\right) $ are the Matsubara frequencies and 
$\theta $ is the angle between the ${\bf x}${\bf \ }axis and the direction
of the Fermi velocity $v_{F}$, and $\tau $ is the elastic scattering time.
In the following, we consider the\ behavior of S/F/S systems close to the
critical transition temperature $T_{c}$, so we may linearize equations $%
\left( \ref{eilenberger complet}\right) $ by putting $g=sign\left( \omega
\right) .$

Usually the electron scattering mean free path in S/F/S systems is rather
small. In such a dirty limit the angular dependence of the Green's functions
is weak, and the Eilenberger equations can be replaced by much simpler
Usadel equations \cite{Usadel70}. In fact, the conditions of the
applicability of Usadel equations are $T_{c}\tau \ll 1$ and $h\tau \ll 1.$
The second condition, is much more restrictive due to a large value of the
exchange field ($h\gg T_{c}$). Therefore the attempts to retain in the
Usadel equations the first correction in the parameter $h\tau $ have been
made in $\cite{Proshin,ProshinReview,baladie}$. They\ have resulted in the
renormalization of the diffusion constant in F-layer $D_{f}\rightarrow
D_{f}(1-2ih\tau ),$ where $D_{f}=v_{F}^{2}\tau /3.$ The critical analysis of
this renormalization in \cite{Fominov} has revealed its inaccuracy, but has
not provided the right correction. To resolve this controversy we perform
below a careful derivation of the Usadel equation for an F-layer retaining
the linear correction over\ parameter $h\tau .$

In the ferromagnet the Cooper pairing is absent, so $\Delta =0$, and
considering the case of the strong exchange field$\ \omega \ll h$, we may
write the linearized Eilenberger equation for a function $f$ as 
\begin{equation}
\left( ih+\frac{1}{2\tau }sign\left( \omega \right) \right) f+\frac{1}{2}%
v_{F}\cos \theta \frac{\partial f}{\partial x}=\frac{1}{2\tau }sign\left(
\omega \right) F,
\end{equation}
where $v_{F}$ and $\tau $ refer to the Fermi velocity and the scattering
time in the ferromagnet. Let us recall that $F$\ is an anomalous Green
function averaged over the angles. Then we may present $f$ as $%
f=F(x)+f_{1}(x,\theta ),$ where $f_{1}\ll F$. So for $\omega >0$%
\begin{equation}
ihf+\frac{f_{1}}{2\tau }+\frac{1}{2}v_{F}\cos \theta \frac{\partial f}{%
\partial x}=0.  \label{eilenbregr}
\end{equation}
The averaging of this equation over the angles gives 
\begin{equation}
ihF+\frac{1}{2}v_{F}\left\langle \cos \theta \frac{\partial f_{1}}{\partial x%
}\right\rangle =0.  \label{eil1}
\end{equation}
Multiplying $\left( \ref{eilenbregr}\right) $ by cos$\theta $, averaging and
taking the derivative over $x$\ we obtain 
\begin{equation}
\left\langle \cos \theta \frac{\partial f_{1}}{\partial x}\right\rangle =-%
\frac{\tau v_{F}}{\left( 2ih\tau +1\right) }\left( \left\langle \cos
^{2}\theta \frac{\partial ^{2}f_{1}}{\partial x^{2}}\right\rangle +\frac{1}{3%
}\frac{\partial ^{2}F}{\partial x^{2}}\right) .
\end{equation}
Then combining $\left( \ref{eilenbregr}\right) $ and $\left( \ref{eil1}%
\right) $ we have 
\begin{equation}
ihF-\frac{\tau v^{2}}{2\left( 2ih\tau +1\right) }\left( \left\langle \cos
^{2}\theta \frac{\partial ^{2}f_{1}}{\partial x^{2}}\right\rangle +\frac{1}{3%
}\frac{\partial ^{2}F}{\partial x^{2}}\right) =0.
\end{equation}
Neglecting in this equation the $f_{1}$ term, we obtain the usual Usadel
equation with the renormalized diffusion constant $D_{f}\rightarrow
D_{f}(1-2ih\tau ).$ However the omitted term gives a correction of the same
order as that coming from the diffusion constant renormalization \cite
{Fominov}. It effectively means that it is needed to keep higher order terms
in the expansion of $f$ over $\cos \theta $ (Legendre polynomials).
Multiplying $\left( \ref{eilenbregr}\right) $ consecutively by $\cos
^{2}\theta $ and $\cos ^{3}\theta $ and taking derivative over $x$, we
obtain after some algebra the following exact equation 
\begin{equation}
ihF-\frac{\tau v_{F}^{2}}{2\left( 2ih\tau +1\right) }\left\{ \frac{-1}{%
\left( 2ih\tau +1\right) }\left[ \frac{2ih\tau }{3}\frac{\partial ^{2}F}{%
\partial x^{2}}-\frac{v_{F}^{2}\tau ^{2}}{\left( 2ih\tau +1\right) }\left(
\left\langle \cos ^{4}\theta \frac{\partial ^{4}f_{1}}{\partial x^{4}}%
\right\rangle +\frac{1}{5}\frac{\partial ^{4}F}{\partial x^{4}}\right) %
\right] +\frac{1}{3}\frac{\partial ^{2}F}{\partial x^{2}}\right\} =0.
\end{equation}
Here we may already safely neglect the $f_{1}$ term, as it gives a
contribution $\sim \left( h\tau \right) ^{2}$. In the result, in the linear
approximation over $h\tau ,$ the Usadel equation is written as 
\begin{equation}
ihF-\frac{\tau v_{F}^{2}}{6\left( 4ih\tau +1\right) }\frac{\partial ^{2}F}{%
\partial x^{2}}-\frac{\tau ^{3}v_{F}^{4}}{10}\frac{\partial ^{4}F}{\partial
x^{4}}=0.  \label{UsadelF1}
\end{equation}
Taking in mind that in zero approximation over $h\tau ,$ the function $F$
satisfies the equation 
\begin{equation}
ihF-\frac{\tau v_{F}^{2}}{6}\frac{\partial ^{2}F}{\partial x^{2}}=0,
\end{equation}
we may finally rewrite $\left( \ref{UsadelF1}\right) $ in the following form
applicable for positive and negative $\omega $%
\begin{equation}
isign\left( \omega \right) hF-\frac{D_{f}\left[ 1-i(2/5)h\tau sign\left(
\omega \right) \right] }{2}\frac{\partial ^{2}F}{\partial x^{2}}=0.
\label{Usadel eq in F}
\end{equation}
So we conclude that the first correction in the parameter $h\tau $ to the
Usadel equation leads to the somewhat different renormalization of the
diffusion constant $D_{f}\rightarrow D_{f}(1-i(2/5)h\tau sign\left( \omega
\right) )$, comparing to what has been suggested before $\cite
{Proshin,ProshinReview,baladie}$ \ $\left( D_{f}\rightarrow D_{f}\left[
1-i2h\tau sign\left( \omega \right) \right] \right) $. This result has been
also independently obtained by Tagirov \cite{Tagirov}.

In the superconducting layer, the linearized Usadel equation is 
\begin{equation}
\left| \omega \right| F(x,\omega )-\frac{D_{s}}{2}\frac{\partial
^{2}F(x,\omega )}{\partial x^{2}}=\Delta (x),  \label{Usadel2}
\end{equation}
\begin{equation}
\Delta (x)=\left| \lambda \right| \pi T\sum_{\omega }F(x,\omega ),
\end{equation}
where $\lambda $ is the BCS coupling constant$\ $and $D_{s}$ is the
diffusion coefficient in S layer.

The equations $\left( \ref{Usadel eq in F}\right) $\ and $\left( \ref
{Usadel2}\right) $ are completed by the general boundaries conditions at the
S/F interfaces \cite{Kuprianov88} 
\[
F_{s}\left( d/2\right) =F_{f}\left( d/2\right) +\xi _{f}\gamma _{B2}\left( 
\frac{\partial F_{f}}{\partial x}\right) _{d/2},
\]
\begin{equation}
F_{s}\left( -d/2\right) =F_{f}\left( -d/2\right) -\xi _{f}\gamma _{B1}\left( 
\frac{\partial F_{f}}{\partial x}\right) _{-d/2},  \label{boundary cond}
\end{equation}
\[
\left( \frac{\partial F_{s}}{\partial x}\right) _{\pm d/2}=\frac{\sigma _{n}%
}{\sigma _{s}}\left( \frac{\partial F_{f}}{\partial x}\right) _{\pm d/2},
\]
where the notation $F_{s}(F_{f})$ is used for the anomalous Green function
in a superconductor (ferromagnet) and $\sigma _{n}$ $\left( \sigma
_{s}\right) $ is the conductivity of the F-layer $\left( \text{S-layer above 
}T_{c}\right) ,$ $\xi _{f}=\sqrt{\frac{D_{f}}{h}\text{ }}$is the
characteristic length of superconducting correlations decaying in F-layer,
and $\xi _{s}=\sqrt{\frac{D_{s}}{2T_{c}}}$ is the superconducting coherence
length of the S-layer, the parameter $\gamma _{B}=\frac{R_{b}\sigma _{f}}{%
\xi _{f}},$ where $R_{b}$ is the S/F boundary resistance per unit area. Here
we introduce two different parameter $\gamma _{B1}$ \ and $\gamma _{B2}$,
for the left and right S/F interfaces, assuming that their boundary
resistances may be different. Note that the parameter $\gamma _{B}$ is
directly related to the transparency of the S/F interface $T=\frac{1}{%
1+\gamma _{B}}$ \cite{AArts}. The limit $T=0$ $\left( \gamma _{B}=\infty
\right) $ corresponds to a vanishingly small boundary transparency, and the
limit $T=1$ $\left( \gamma _{B}=0\right) $ corresponds to a perfectly
transparent interface.

In principle, the equations $\left( \ref{Usadel eq in F}\right) $ and $%
\left( \ref{Usadel2}\right) $ with boundary conditions $\left( \ref{boundary
cond}\right) $ give the complete description of the S/F/S junction near the
transition temperature.

\section{Calculation of the Josephson critical current}

Let us start with a general solution of the Usadel equation in the
ferromagnet for $\omega >0$%
\begin{equation}
F_{f}\left( \omega >0,x\right) =A_{\omega }\sinh (kx)+B_{\omega }\cosh (kx),
\end{equation}
where $k=\left[ \left( 1-0.2h\tau \right) +i(1+0.2h\tau \right] \sqrt{\frac{h%
}{D_{f}}}.$ Similarly for $\omega <0$ 
\begin{equation}
F_{f}\left( \omega <0,x\right) =C_{\omega }\sinh (k^{\ast }x)+D_{\omega
}\cosh (k^{\ast }x)
\end{equation}

We see that the complex wave-vector $k$ describes the oscillating
exponential damping of the anomalous Green function inside the ferromagnet.
In the case of very small scattering time $\left( \tau \rightarrow 0\right) ,
$ the imaginary and real parts of the wave-vector $k$ coincide, and so the
characteristic lengths for oscillating and damping are exactly the same. The
finite scattering length decreases the oscillating period and increases the
damping length. This is consistent with the fact that in a pure limit the
exponential damping is replaced by a much slower power damping, while the
oscillations have a much shorter period comparing to the dirty limit \cite
{Panyukov82,baladie,volkov}.

To obtain the analytical description of the critical current in S/F/S
junctions, it is useful to introduce the functions $F^{+}$ and $F^{-}$ \cite
{Kuprianov91}
\begin{eqnarray}
F^{+}\left( \omega >0,x\right)  &=&F\left( \omega >0,x\right) +F\left(
\omega <0,x\right) , \\
F^{-}\left( \omega >0,x\right)  &=&F\left( \omega >0,x\right) -F\left(
\omega <0,x\right) .
\end{eqnarray}
In a superconductor, the equation for $F_{s}^{+}$ is the same as $\left( \ref
{Usadel2}\right) $ with $\Delta \rightarrow 2\Delta $, while the equation
for $F_{s}^{-}$ is much simpler 
\begin{equation}
\left| \omega \right| F_{s}^{-}(x,\omega )-\frac{D_{s}}{2}\frac{\partial
^{2}F_{s}^{-}(x,\omega )}{\partial x^{2}}=0,  \label{fs-}
\end{equation}
and the\ self consistency equation reads as 
\begin{equation}
\Delta =\left| \lambda \right| \pi T\sum_{\omega >0}F_{s}^{+}(x,\omega ).
\end{equation}
The boundary conditions $\left( \ref{boundary cond}\right) $ will be the
same for $F^{+}$ and $F^{-}$ functions too. Using the solution of $\left( 
\ref{fs-}\right) $ for $F_{s}^{-}(x,\omega )$, it may be demonstrated\ (see
also\cite{Kuprianov91}) that in the case of a rather large resistivity of
the F layer $\left( \frac{\sigma _{n}}{\sigma _{s}}\frac{\xi s}{\xi _{f}}\ll
1\right) $ or$\ $in the case low transparency of the interfaces, when the
condition $\frac{\sigma _{n}}{\sigma _{s}}\frac{\xi s}{\xi _{f}}\frac{1}{%
\gamma _{_{B}}}\ll 1$ is satisfied, the boundary conditions for $F_{f}^{-}$
functions are essentially simplified
\begin{eqnarray}
F_{f}^{-}\left( d/2\right) +\xi _{f}\gamma _{B2}\left( \frac{\partial
F_{f}^{-}}{\partial x}\right) _{_{d/2}} &=&0, \\
F_{f}^{-}\left( -d/2\right) -\xi _{f}\gamma _{B1}\left( \frac{\partial
F_{f}^{-}}{\partial x}\right) _{_{-d/2}} &=&0.
\end{eqnarray}
Further on, introducing the functions 
\begin{eqnarray}
{\cal F}^{+} &=&\left| \lambda \right| \pi T\sum_{\omega >0}F^{+}\left(
\omega >0,x\right) , \\
{\cal F}^{-} &=&\left| \lambda \right| \pi T\sum_{\omega >0}F^{-}\left(
\omega >0,x\right) ,
\end{eqnarray}
we obtain the following system of equations for the order parameter and its
derivatives on the left and right sides of the junction 
\begin{gather}
{\cal F}_{f}^{-}\left( d/2\right) +\xi _{f}\gamma _{B2}\left( \frac{\partial 
{\cal F}_{f}^{-}}{\partial x}\right) _{d/2}=0, \\
{\cal F}_{f}^{-}\left( -d/2\right) -\xi _{f}\gamma _{B1}\left( \frac{%
\partial {\cal F}_{f}^{-}}{\partial x}\right) _{-d/2}=0, \\
\Delta \left( x=d/2\right) =\Delta ^{+}={\cal F}_{f}^{+}\left( d/2\right)
+\xi _{f}\gamma _{B2}\left( \frac{\partial {\cal F}_{f}^{+}}{\partial x}%
\right) _{d/2}, \\
\Delta \left( x=-d/2\right) =\Delta ^{-}={\cal F}_{f}^{+}\left( -d/2\right)
-\xi _{f}\gamma _{B1}\left( \frac{\partial {\cal F}_{f}^{+}}{\partial x}%
\right) _{-d/2}, \\
\left( \frac{\partial \Delta ^{+}}{\partial x}\right) =\frac{\sigma _{n}}{%
\sigma _{s}}\left( \frac{\partial {\cal F}_{f}^{+}}{\partial x}\right)
_{d/2}, \\
\left( \frac{\partial \Delta ^{-}}{\partial x}\right) =\frac{\sigma _{n}}{%
\sigma _{s}}\left( \frac{\partial {\cal F}_{f}^{+}}{\partial x}\right)
_{-d/2}.
\end{gather}
This system of six equations, after the elimination of four coefficients\ of
the type $\left( \left| \lambda \right| \pi T%
\mathrel{\mathop{\sum }\limits_{\omega >0}}%
A_{\omega }\right) $ may be reduced to a standard linear form of $\left( 
\frac{\partial \Delta ^{+}}{\partial x}\right) ,$ $\left( \frac{\partial
\Delta ^{-}}{\partial x}\right) $, $\Delta ^{+}$ and $\Delta ^{-}$%
\begin{eqnarray}
\Delta ^{+} &=&M_{11}\Delta ^{-}+M_{12}\left( \frac{\partial \Delta ^{-}}{%
\partial x}\right) , \\
\left( \frac{\partial \Delta ^{+}}{\partial x}\right)  &=&M_{21}\Delta
^{-}+M_{22}\left( \frac{\partial \Delta ^{-}}{\partial x}\right) ,
\end{eqnarray}
which was used by De Gennes in his general description of the Josephson
junction close to $T_{c}$ \cite{De Gennes}$.$ Following the approach \cite
{De Gennes}$,$ the critical current of the SFS junction is given directly by
the expression 
\begin{equation}
I_{c}=\frac{-ie\pi N_{s}(0)D_{s}}{4T_{c}M_{12}}\left( \Delta ^{\ast -}\Delta
^{+}-c.c\right) ,
\end{equation}
where $N_{s}(0)$ is the electron density of states in S electrodes when they
are in the normal state.

The calculation of the coefficient $M_{12}$ is straightforward but rather
cumbersome for a general case. The corresponding formula is to intricate to
be presented here. So, to discuss the essential physics, we consider below
only the most interesting limits.

In the case of highly transparent interfaces, it is possible to obtain an
analytical expression of the critical current taking into account the first
order correction over $h\tau $ in the Usadel equation $\left( \ref{Usadel eq
in F}\right) $%
\begin{equation}
I_{c}=\frac{\pi \Delta ^{2}}{8eT_{c}}\frac{\sigma _{n}}{\xi _{f}}\left| 
\frac{k}{\sinh \left( \frac{kd}{2}\right) \cosh \left( \frac{kd}{2}\right) }%
+c.c.\right| 
\end{equation}
where the complex wave vector $k=\left[ \left( 1-0.2h\tau \right)
+i(1+0.2h\tau \right] \sqrt{\frac{h}{D_{f}}}.$ The dependences $I_{c}(d)$
for different parameter $h\tau $ are presented in Fig. $2$.

When the critical current goes through zero, the transition from $0$ to $\pi 
$-shift Josephson contact takes place. As expected, with the increase of the
scattering time the period of oscillations becomes somewhat shorter, while
its amplitude increases. This is consistent with a power low decrease of the
critical current with F-layer thickness in the clean limit \cite{Panyukov82}.

Further we concentrate on the influence of the S/F boundary transparency on
the $I_{c}$ oscillations, and then we neglect the $h\tau $ correction in the
diffusion coefficient. Starting with the case of completely transparent
interfaces $\gamma _{B1,2}\rightarrow 0,$ we retrieve the corresponding
expression previously obtained in \cite{Kuprianov91}
\begin{equation}
I_{c}=\frac{\sigma _{n}}{\xi _{f}}\frac{\pi \Delta ^{2}}{eT_{c}}\left| \frac{%
\sin (x)\cosh (x)+\cos (x)\sinh (x)}{\cos (2x)-\cosh (2x)}\right| ,
\end{equation}
where $x=d/\xi _{f}.$ In the case of low transparency of both interfaces $%
\gamma _{B1,2}\gg 1,$ the critical current is 
\begin{equation}
I_{c}=\frac{1}{\gamma _{B1}\gamma _{B2}}\frac{\sigma _{n}}{\xi _{f}}\frac{%
\pi \Delta ^{2}}{2eT_{c}}\left| \frac{\cos (x)\sinh (x)-\sin (x)\cosh (x)}{%
\cos (2x)-\cosh (2x)}\right| .
\end{equation}
If one interface (the left S/F interface) is transparent, $\gamma
_{B1}\rightarrow 0,$ the critical current is 
\begin{equation}
I_{c}=\frac{\sigma _{n}}{\xi _{f}}\frac{\pi \Delta ^{2}}{eT_{c}}\left| \frac{%
\cosh (x)\sin (x)+\cos (x)\sinh (x)+2\gamma _{B2}\cos (x)\cosh (x)}{%
(1-2\gamma _{B2}^{2})\cos (2x)-(1+2\gamma _{B2}^{2})\cosh (2x)-2\gamma _{B2}%
\left[ \sin (2x)+\sinh (2x)\right] }\right|   \label{gamma=zero}
\end{equation}
If one interface (the first S/F interface) has a low transparency, $\gamma
_{B1}\gg 1,$ the critical current is 
\begin{equation}
I_{c}=\frac{\sigma _{n}}{\gamma _{B1}\xi _{f}}\frac{\pi \Delta ^{2}}{eT_{c}}%
\left| \frac{\cos (x)\cosh (x)-\gamma _{B2}\left[ \cosh (x)\sin (x)-\cos
(x)\sinh (x)\right] }{(1-2\gamma _{B2}^{2})\cos (2x)+(1+2\gamma
_{B2}^{2})\cosh (2x)-2\gamma _{B2}\left[ \sin (2x)-\sinh (2x)\right] }%
\right|   \label{gamma=inf}
\end{equation}

The evolution of the $I_{c}(d)$ dependence with the decrease of the
transparency of the second interface is presented in Fig. $3$. We observe
that with the increase of $\gamma _{B2}$ \ the amplitude of oscillations
decreases, as well as the thickness of the ferromagnetic layer corresponding
to the first zero of the critical current.

In the experiment \cite{Kontos2}, the presence of Al$_{2}$O$_{3}$ barrier at
one S/F boundary can be modeled by a low transparency interface $\left(
\gamma _{B1}\gg 1\right) $ while the other boundary is quite transparent $%
\gamma _{B2}\rightarrow 0.$ The resistance $R_{n}$ per unit area of the
S/F/S junction in this experiment is dominated by the resistance of the
tunnel barrier and thus $R_{n}\sim \frac{\gamma _{B1}\xi _{f}}{\sigma _{n}}$%
. So we may expect that the experimental situation \cite{Kontos2} must be
described by the expression $\left( \ref{gamma=inf}\right) $ with $\gamma
_{B2}=0$ i.e.
\begin{equation}
I_{c}R_{n}=\frac{\pi \Delta ^{2}}{eT_{c}}\left| \frac{\cos (x)\cosh (x)}{%
\cos (2x)+\cosh (2x)}\right| .  \label{fit}
\end{equation}
Note that this expression is quite different from that used in \cite{Kontos2}
to fit the experimental data. Unfortunately in \cite{Kontos2} there is no
information which could shed light not only on the reasons of the
discrepancy between our formula but even at the origin of the used
theoretical expression. The experimental results on the $I_{c}(d)$
oscillations presented in \cite{Kontos2} have been obtained at low
temperature. Nevertheless we think that the overall shape of the $I_{c}(d)$
curve is not very sensitive to the temperature and we compare the
experimental data \cite{Kontos2} with our theoretical expression $\left( \ref
{fit}\right) $ in Fig. $4.$ We use as the fitting parameter $\xi _{f}\sim 30$
\AA , while the experiment \cite{Kontos1} provided a value of $\xi _{f}$
around $35$ \AA\ in our notation. The value of the other fitting parameter $%
\frac{\pi \Delta ^{2}}{eT_{c}}\ $is $110$ $\mu V$; this parameter is quite
difficult to estimate because of the uncertainty on the value of $\Delta $
at the S/F interface in the geometry of experiment \cite{Kontos2}$.$ Also
following the analysis of the authors \cite{Kontos2} \ presented in their
previous publication\cite{Kontos1}, we have taken into account that the
actual ferromagnetic thickness of PdNi layer is reduced by $15$ \AA\ due to
some interdiffusion at the S/F interface. The obtained description of the
experimental data \cite{Kontos2} is quite satisfactory.

In conclusion, we have presented the general theoretical description of the
ferromagnetic $\pi $-junctions near superconducting transition temperature
and proposed a simple way to take into account the finite elastic scattering
time in Usadel equation. The obtained analytical expressions may be useful
for the analysis of different experimental realizations of such junctions.
Our analysis may be also easily generalized to the situation when the
superconducting electrodes are fabricated from different materials.

We thank M. Aprili and V. Ryazanov for stimulating discussions. This work
was supported by the ESF ``vortex'' Programme and the ACI
``supra-nanom\'{e}trique''.

FIG. $1$ Geometry of the considered S/F/S system. The thickness of the
ferromagnetic layer is $d$. The transparency of the left S/F interface is
characterized by the coefficient $\gamma _{B1}$ and the transparency of the
right F/S interface is characterized by $\gamma _{B2}.$

FIG. $2$ Critical current of the S/F/S junction as a function of the
thickness of the ferromagnetic layer normalized by $\xi _{f}$ for different
scattering time. Both S/F interfaces are completely transparent. The
parameter $h\tau $ is equal to $0$, $0.25$ and $0.5$.

FIG. $3$ Critical current of the S/F/S junction as a function of the
thickness of the ferromagnetic layer normalized by $\xi _{f}$. The first S/F
interface has a low transparency $\left( \gamma _{B1}\gg 1\right) $. The
parameter $\gamma _{B2}$ characterizing the transparency of the second
interface is chosen as $0.2,$ $1.5,$ $3.0,$ and $10.$

FIG. $4$ The experimental points correspond to the measurement of the
critical current, done by Kontos et al \cite{Kontos2}, vs the PdNi layer
thickness. The theoretical curve is the best fit obtained by using formula $%
\left( \ref{fit}\right) $. The fitting parameters are $\xi _{f}\sim 30$ \AA\ %
and $\frac{\pi \Delta ^{2}}{eT_{c}}\sim 110$ $\mu V.$

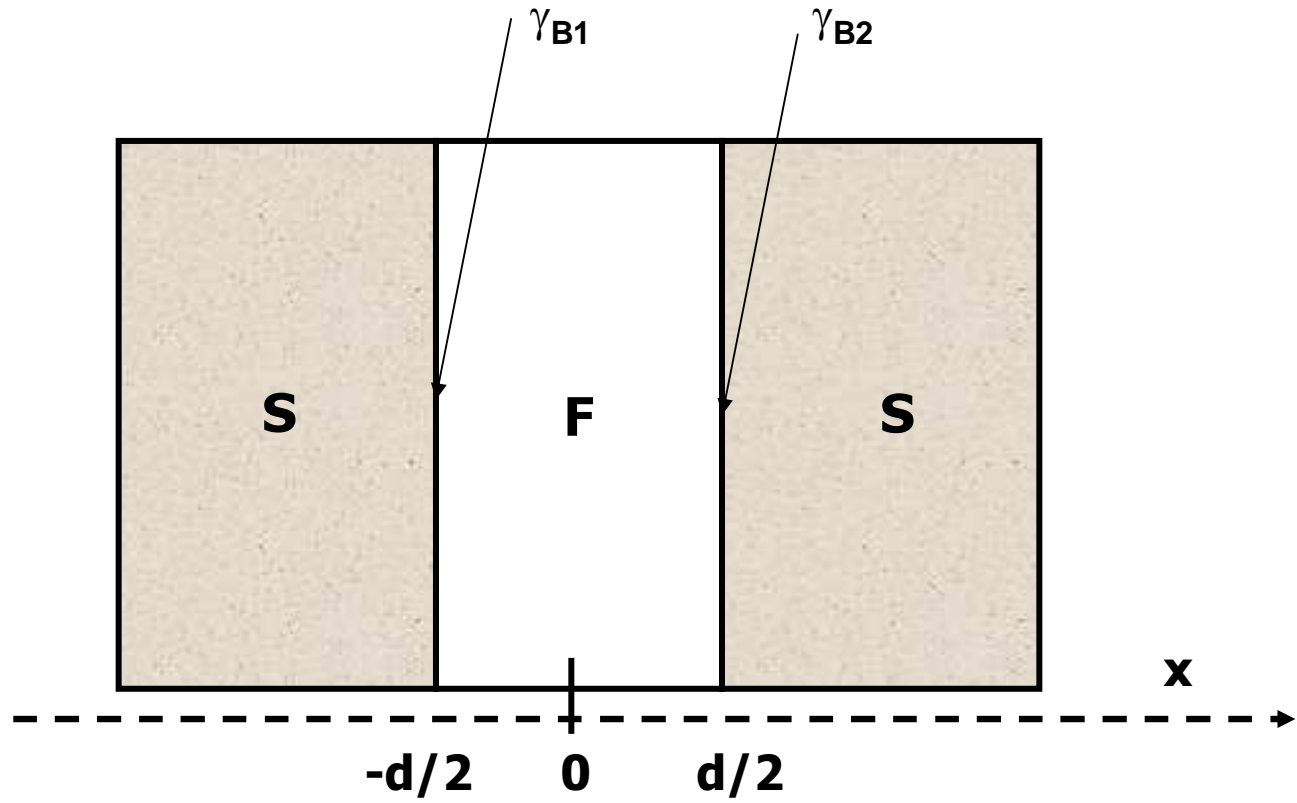
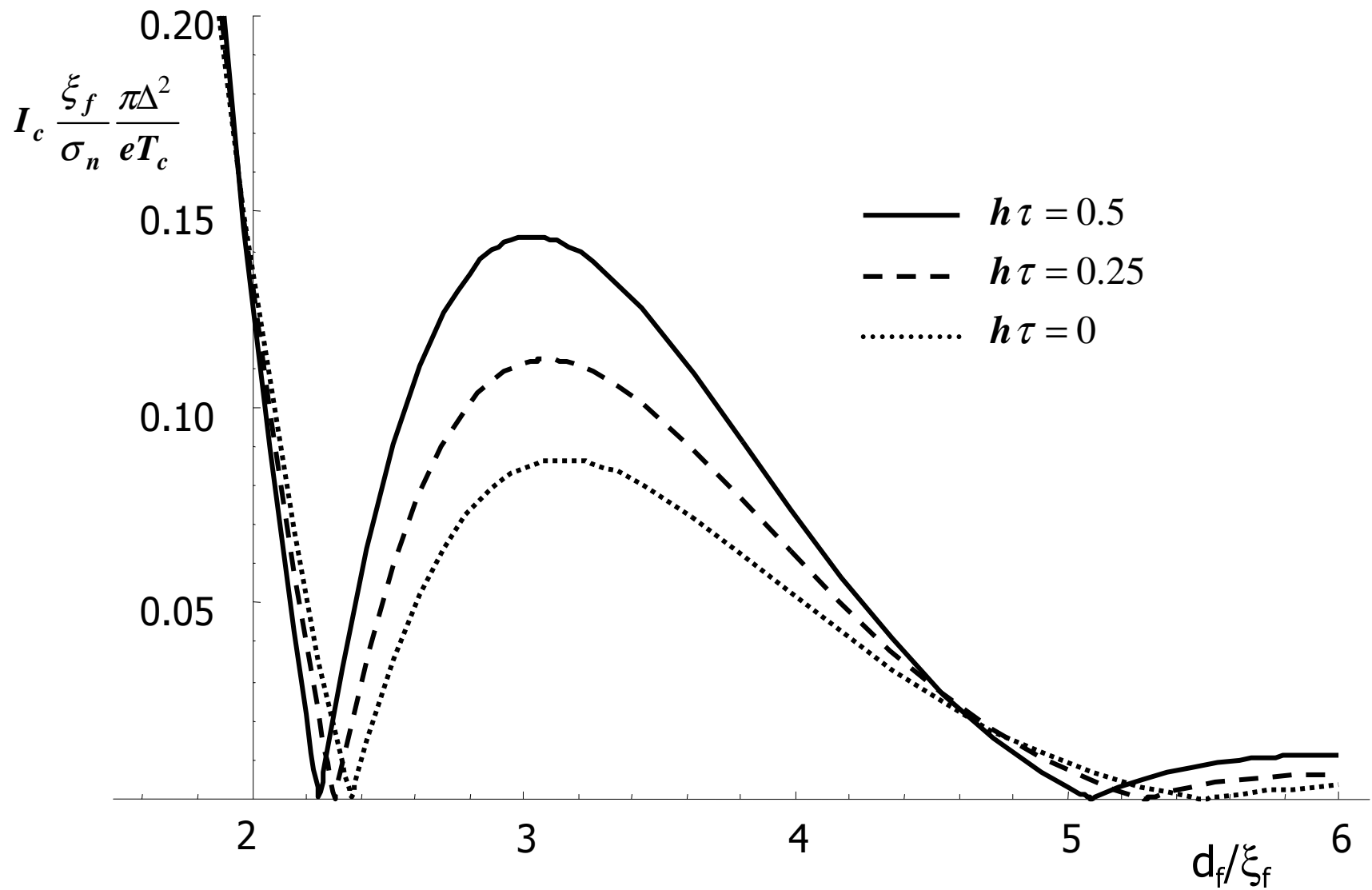
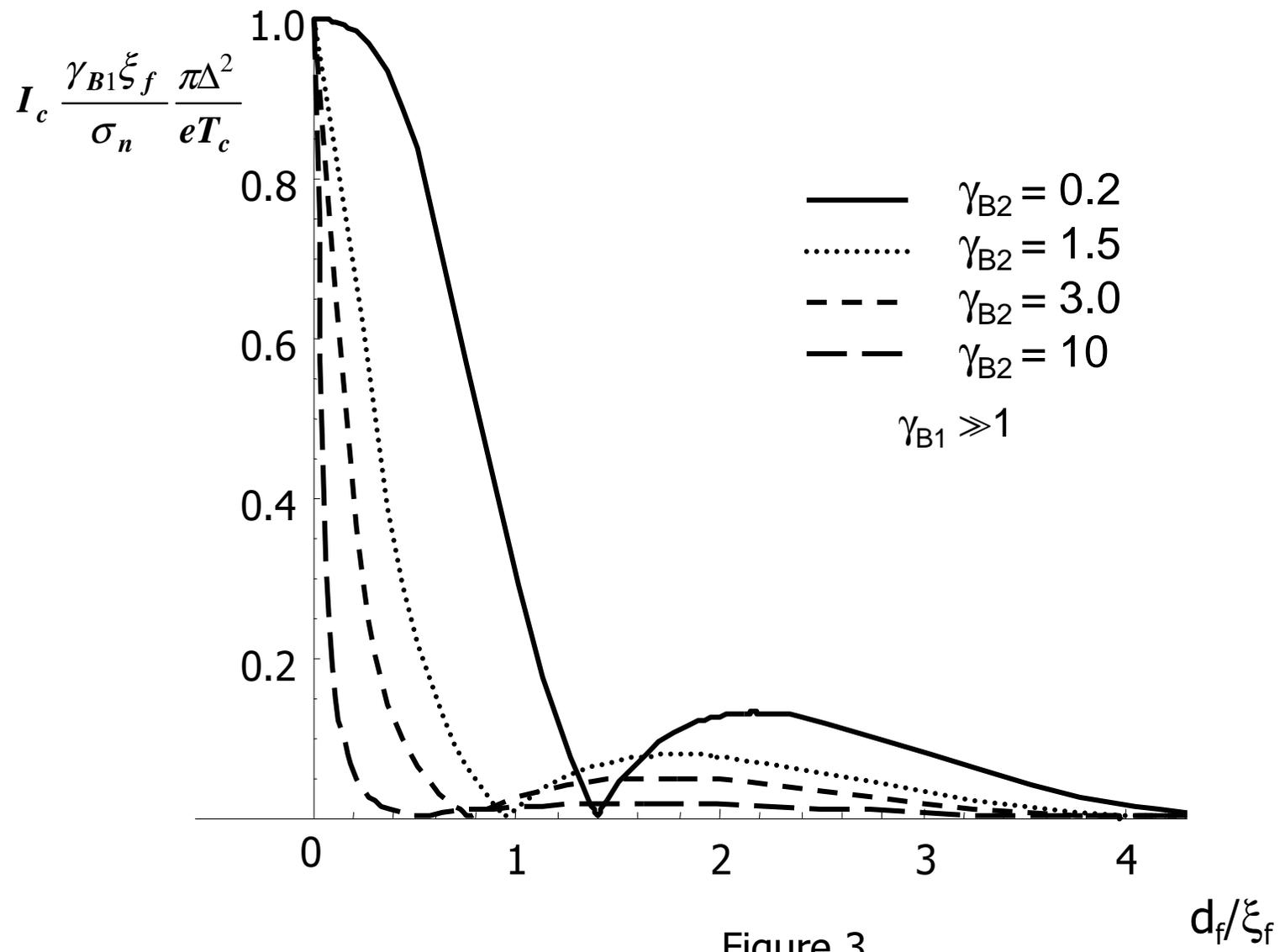
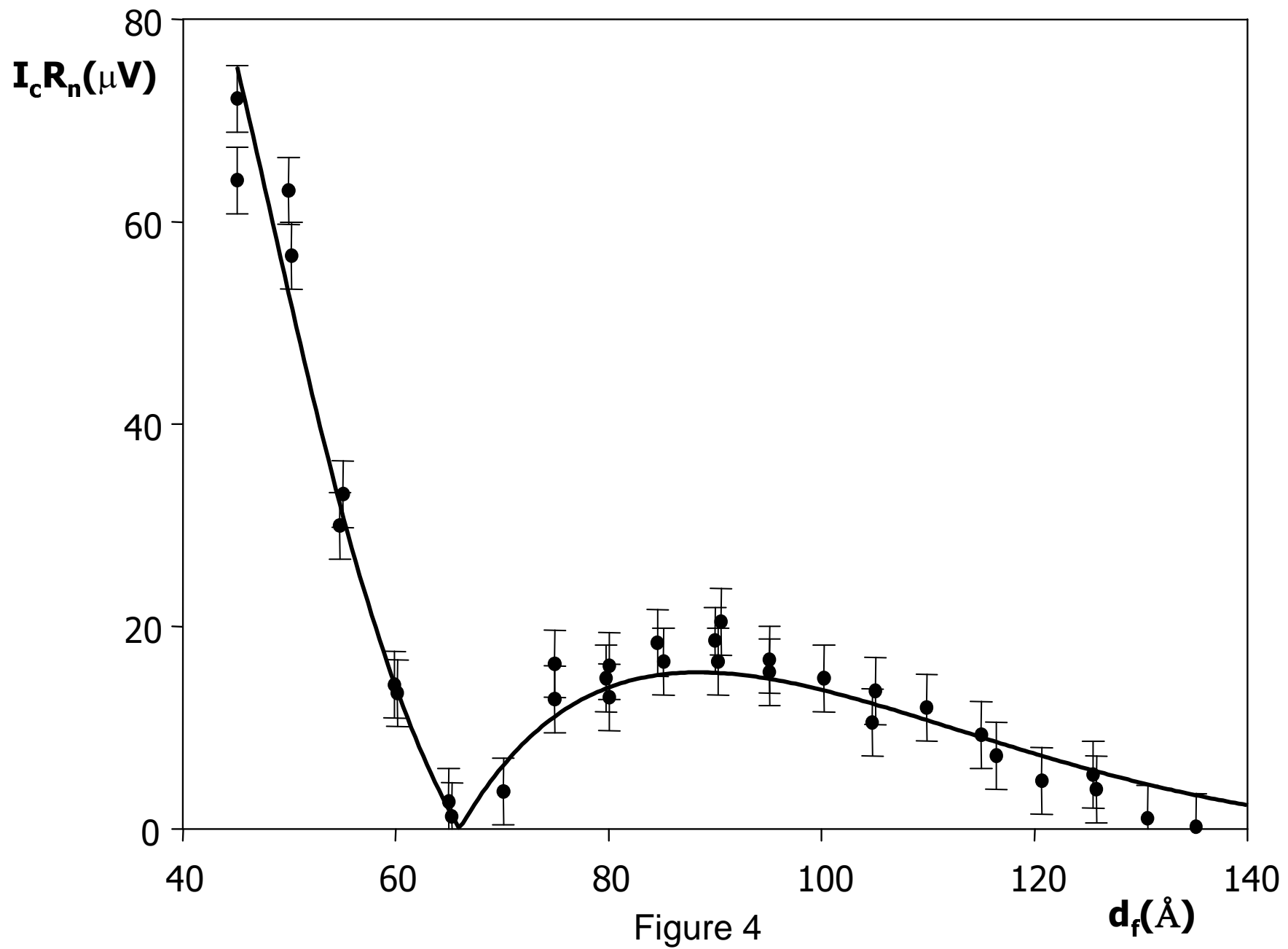
\end{document}